\documentclass[aps,prb,twocolumn,superscriptaddress]{revtex4}
\usepackage{amsmath}
\usepackage{amssymb}
\usepackage{graphicx}

\usepackage{color}
\usepackage{subfigure}
\usepackage{bm}
\usepackage{epstopdf}

\graphicspath{{figures}}

\begin{document}

\title{Spin transport in Aharonov-Bohm ring with exchange interaction}

\author{I. G. Savenko}
\affiliation{Science Institute, University of Iceland, Dunhagi 3,
IS-107, Reykjavik, Iceland}
\affiliation{Department of Applied Physics/COMP, Aalto University, PO Box 14100, 00076 Aalto, Finland}

\author{R. G. Polozkov}
\affiliation{St.-Petersburg State Polytechnical University, Politechnicheskaya 29, 195251 St.Petersburg, Russia}

\author{I. A. Shelykh}
\affiliation{Science Institute, University of Iceland, Dunhagi 3,
IS-107, Reykjavik, Iceland}
\affiliation{Division of Physics and Applied Physics, Nanyang Technological University 637371, Singapore}

\date{\today}

\begin{abstract}
We investigate spin-dependent conductance through a quantum Aharonov-Bohm ring containing localized electrons which interact with the propagating flow of electrons via exchange interaction of the ferromagnetic or antiferromagnetic type. We analyze the conductance oscillations as a function of both the chemical potential determined by the concentration of the carriers and external magnetic field. It is demonstrated that the amplitude of the conductance oscillations in the ballistic regime is governed by the value of the non-compensated spin localized in the ring. The results are in agreement with the concept of fractional quantization of the ballistic conductance, proposed by us earlier [Phys. Rev. B 71,113311 (2005)].
\end{abstract}

\pacs{73.23.Ad, 73.21.Hb, 73.63.Nm, 72.25.Dc, 85.35.Ds, 85.75.Hh}

\maketitle

%-----------------------------------------------------------------
%-----------------------------------------------------------------
%-----------------------------------------------------------------

\section{Introduction}

Recent progress in nanotechnology has allowed to prepare quasi one-dimensional (1D) semiconductor systems containing high-mobility charge carriers, which exhibit ballistic behavior in the regime when the decoherence time is longer than the characteristic time needed for an electron or a hole to pass through the structure. In this case the transport of charge carriers in such systems is of coherent nature. Since the ballistic transport is not accompanied by the Joule losses, the conductance of the quasi 1D semiconductor systems in a single mode regime at small drain-source biases depends only on the transmission coefficient describing the elastic scattering in the ballistic region~\cite{Pepper1986,Wharam1988}. The latter is determined by the geometry of the system and can depend on such parameters, as the Fermi energy of the carriers and external electric or magnetic fields.

In some cases, the scattering becomes spin-dependent. This happens if the ballistic region contains confined electrons possessing  non-compensated spin which interacts with the spins of the conducting electrons. A standard textbook example of this phenomenon is Kondo related phenomena in transport through individual quantum dots connected to one dimensional leads \cite{GoldhaberPRL,GoldhaberNature}. Another example is the formation of the ``0.7 anomaly'' in the ballistic conductance of an individual quantum point contact (QPC) split off from the first step in the quantum conductance staircase \cite{Thomas1996,Bagraev2002,Reilly2002}. Although the exact mechanism of the formation of the ``0.7 anomaly'' is still a matter of debates~\cite{ShelykhJPCM}, several experimental observations has indicated the importance of the spin component for the behavior of this feature. First, an electron g-factor is found to raise from 0.4 to 1.3 as the number of occupied 1D subbands decreases~\cite{Thomas1996}. Second, the height of the feature attains to a value of $0.5$ in a strong external magnetic field \cite{Thomas1998,Thomas2000}.

These results have defined the spontaneous spin polarization of a 1D gas in a zero magnetic field as one of the possible mechanisms of the appearence of the feature~\cite{Chuan1998,Starikov2003,Bagraev2004,Rejec2006,Rokhinson2006,Lind2011}. The localized spins affect the propagating carriers via exchange interaction of either ferromagnetic or antiferromagnetic type. Since it is defined by the mutual orientation of spins, the transmission coefficient through the QPC becomes spin-dependent. If the QPC contains a single localized electron, the eigenstates of the system consisting of localized and transmitted electrons, are the essence of singlet and triplet states \cite{Flambaum2000,Rejec2003}. If the energy of the triplet state is lower than the one of the singlet state, the potential barrier for the carriers in the singlet configuration becomes higher than one for the triplet state. Therefore, at small concentrations of carriers, the ingoing electron in the triplet configuration passes by the ring region freely, while the carriers in the singlet configuration are reflected. In zero magnetic field the probability of realization of the triplet configuration equals to $3/4$ against $1/4$ in the case of the singlet one; thus the full conductance in the considered regime equals $G = 0.75(2e^2/h)$ ~\cite{Flambaum2000}. In contrast, if the singlet configuration is energetically preferable, the conductance equals to $G = 0.25(2e^2/h)$ \cite{Rejec2000,Rejec2003}. If the uncompensated spin of electrons localized in the QPC $J>1/2$, the quantization pattern becomes more complicated, and the value of the fractional plateau becomes $G = (J+1)/(2J+1)(2e^2/h)$ for the ferromagnetic interaction and $G = (J)/(2J+1)(2e^2/h)$ for the antiferromagnetic interaction \cite{Shelykh2006}.

Spin related effects become even more pronounced in the ballistic transport through non-single connected objects, such as Aharonov-Bohm (AB) rings. Spin-orbit interaction of the Rashba type in the AB ring induces the Aharonov-Casher (AC) and Berry phase shifts between the spin waves propagating in the clockwise and anticlockwise directions, which results in the large conductance modulations due to the interference of the spin wavefunctions \cite{Aronov1993}. Experimental observation of the AC oscillations in the gate-controlled AB rings has been reported for both electrons~\cite{Nitta,Bergstein} and holes~\cite{Bagraev2006}. Moreover, formation of localized states with uncompensated spin can be also expected to affect the transport properties of the rings. It was shown that the insert of a quantum point contact (QPC) in one of the ring's arms, changes the conductance pattern of the AB ring significantly due to the exchange interaction between the electron localized inside the QPC and a propagating electron. It manifests itself in the formation of ``0.7 feature'' on the quantum conductance staircase of the whole ring structure~\cite{Shelykh2007}.

In the current work, we analyse theoretically spin-dependent transport in a double-slit AB ring with confined delocalized spontaneously spin polarized electrons which interact with conductance electrons passing through the ring in the ballistic regime. The localized spins affect the propagating carriers via exchange interaction of either ferromagnetic or antiferromagnetic type. Since it is defined by the mutual orientation of spins, the transmission coefficient through the ring becomes spin-dependent, which is reflected in the dramatic changes of the patterns of the conductance as a function of the Fermi energy and external magnetic field.

The paper is organized as follows. In Section II we describe the model of the device we consider in our work. In Section III we present the formalism for the calculation of the conductance in the ballistic regime and in Section IV give the describe the obtained results. Section V contains conclusions of the work.

%-----------------------------------------------------------------
%-----------------------------------------------------------------
%-----------------------------------------------------------------

\section{The model}

A sketch of the device we consider is presented in Fig.~1. It represents a quantum AB ring containing spin-polarized interacting electron gas connected to two symmetrically placed leads which serve as a source and a drain of electrons from the two opposite sides. Both the ring and the leads are considered to be 1D in order to make semi-analytical treatment possible. This approximation is legitimate as long as the Fermi energy and the leads cross-section area are small enough and the condition that only lowest subband of the dimensional quantization is occupied holds: $mL^2E_F/\pi^2\hbar^2<1$.~\cite{Shelykh2007}. The ring contains spontaneously spin polarized electron gas with uncompensated spin $J\geq1/2$.
\begin{figure}[!b]
\label{FigSketch}
\centering
\includegraphics[width=0.75\linewidth]{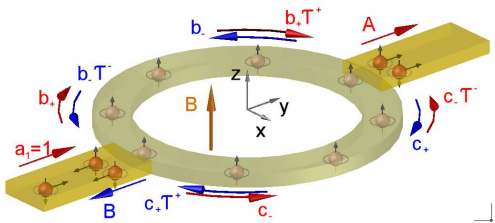}
\caption{Sketch of the system: a double-slit AB ring with localized electrons interacting with the propagating charged carriers via the Coulomb and exchange interaction. The amplitudes of the waves which enter into equations for the determination of transmission and reflection probabilities are shown. $A$ and $B$ correspond to the transmission and reflection amplitudes. Due to the exchange interaction between propagating electrons and spin- polarized electrons in the ring these amplitudes may become spin- dependent as discussed in the main text. }
\end{figure}

The AB ring and the leads are connected by means of two identical QPCs, scattering on which is considered to be of elastic nature. Besides, scattering on them is presumed spin-independent, thus we assume that the spin of a carrier is conserved while passing through the QPCs. Thus, the propagating electrons are supposed to interact with the localized spin only in the region of the ring. The main parameter of a QPC is the amplitude of elastic backscattering of a carrier propagating inside the lead: $\sigma$, $|\sigma| < 1$, determined by the system geometry (a QPC becomes transparent if $\sigma=0$).

In our research we consider zero temperature, hence the charge carriers (electrons) have a step-like distribution in the leads. The drain-source voltage $V_{d}$ is also considered to be small enough, $eV_{d} \ll E_F$.  Thus only those electrons whose energy lies in the vicinity of the Fermi energy can take part in the transport. The radius of the AB ring $R$ is taken much smaller than the inelastic scattering length to ensure the validity of ballistic transport approximation and use of well known Landauer-Buttiker approach to calculate the conductance.~\cite{Landauer1957,Buttiker1986}

An external magnetic field $B$ is applied perpendicularly to the plane of the AB ring. This field influences both the spatial and spin coordinates of the electrons propagating inside the AB ring and the leads and thereby defines the Aharonov-Bohm phase shift and the Zeeman splitting. The latter we neglect for simplicity  of derivation and transparency of results.

Our main goal will be to calculate the dependence of the conductance of the system on chamical potential (Fermi energy) and external magnetic field in various regimes.

%-----------------------------------------------------------------
%-----------------------------------------------------------------
%-----------------------------------------------------------------

\section{The Formalism}

Let us first consider the case when AB ring contains a single localized electron. The full Hamiltonian of the system in the basis of uncoupled states dorresponding to the spins of condictance (\emph{e}) and localized (\emph{S}) electrons ($|\uparrow_e\uparrow_S\rangle$, $|\downarrow_e\downarrow_S\rangle$, $|\uparrow_e\downarrow_S\rangle$, $|\downarrow_e\uparrow_S\rangle$) can be written in the form:
\begin{eqnarray}
\label{EgFullHam}
{\cal H}
%&=&
=
{\cal H}_0+{\cal H}_{int}~~~~~~~~~~~~~~~~~~~~~~~~~~~~~~~~~~~~~~~~~~~~~~~~~~~~\\
\nonumber
%&=&
=
\left(
\begin{array}{cccc}
\frac{\hbar^2k^2}{2m}+V^- & 0 & 0 & 0 \\
0 & \frac{\hbar^2k^2}{2m}+V^- & 0 & 0 \\
0 & 0 & \frac{\hbar^2k^2}{2m}+V^+ & -2V_{ex}\\
0 & 0 & -2V_{ex} & \frac{\hbar^2k^2}{2m}+V^+
\end{array}
\right),
\end{eqnarray}
where interaction of the localized and propagating electrons inside the QPC is modelled
within the framework of the Heisenberg exchange Hamiltonian:
\begin{equation}
{\cal H}_{int} = V_{dir}-V_{ex}~\mathbf{\bm{\sigma}}_e\cdot\mathbf{\bm{\sigma}}_S.
\label{EgDirExHam}
\end{equation}
Here $V_{dir}$ characterizes the Coulomb interaction between the moving and localized electrons plus the effect of the applied bias voltage; $V_{ex}$ corresponds to the exchange interaction; $V^\pm=V_{dir}\pm V_{ex}$; the spin operators $\mathbf{\bm{\sigma}}_e$ and $\mathbf{\bm{\sigma}}_S$ correspond to the propagating and localized electrons, respectively.

Hamiltonian (\ref{EgFullHam}) can be easily diagonalized by the canonical transformation,~\cite{Shelykh2007} and we obtain:
\begin{eqnarray}
\label{EgHamDiag}
{\cal H}=
\left(
\begin{array}{cccc}
\frac{\hbar^2k^2}{2m}+V^- & 0 & 0 & 0 \\
0 & \frac{\hbar^2k^2}{2m}+V^- & 0 & 0 \\
0 & 0 & \frac{\hbar^2k^2}{2m}+V^- & 0\\
0 & 0 & 0 & \frac{\hbar^2k^2}{2m}+V_3
\end{array}
\right),
\end{eqnarray}
where we used the notation: $V_3=V_{dir}-3V_{ex}$.

Hamiltonian ${\cal H}$ corresponds to the new eigensates of the coupled system:
$|1\rangle=|\uparrow_e\uparrow_S\rangle$,
$|2\rangle=|\downarrow_e\downarrow_S\rangle$,
$|3\rangle=\frac{1}{\sqrt{2}}\left(|\uparrow_e\downarrow_S\rangle +
|\downarrow_e\uparrow_S\rangle\right)$,
$|4\rangle=\frac{1}{\sqrt{2}}\left(|\uparrow_e\downarrow_S\rangle -
|\downarrow_e\uparrow_S\rangle\right)$.
The first three states above correspond to the triplet (\emph{t}) configuration, whereas the fourth one is related to the singlet (\emph{s}) configuration.
These modes are characterized by the following wavevectors:
\begin{eqnarray}
\label{Eqk1}
k_{t}=\sqrt{\frac{2m}{\hbar^2}\left[\mu-V_{dir}+V_{ex}\right]},\\
\label{Eqk2}
k_{s}=\sqrt{\frac{2m}{\hbar^2}\left[\mu-V_{dir}-3V_{ex}\right]}.
\end{eqnarray}
In a zero magnetic field, the singlet and triplet state energies are split by the value of $4V_{ex}$.
In the case of ferromagnetic coupling ($V_{ex} > 0$), the triplet configuration has lower energy and thus corresponds to the ground state; the opposite is valid for the antiferromagnetic coupling ($V_{ex} < 0$).

%-----Conductance

Due to the processes of exchange interaction between the localized and freely propagating electrons, the latter can either conserve their spin projection or undergo a spin-flip. The probabilities of these processes depend on the mutual orientation of the two spins as well as on the exchange matrix element, $V_{ex}$.
If the spins of the localized and propagating electrons are parallel, only spin-conservative processes are allowed. On the other hand, if the spins are antiparallel, a spin-flip process becomes possible.
Using the formalism of Ref.\cite{Shelykh2007}, we can write the expression for the ballistic conductance of the AB ring with exchange interaction (at finite temperature):
\begin{widetext}
\begin{eqnarray}
%\nonumber
G\approx\frac{e\hbar}{2\pi m}
\int_0^\infty
\{
\left[
P_{S\uparrow}|A_{e\uparrow S\uparrow\rightarrow e\uparrow S\uparrow}(k)|^2
+P_{S\downarrow}
\left(|A_{e\uparrow S\downarrow\rightarrow e\uparrow S\downarrow}(k)|^2+|A_{e\uparrow S\downarrow\rightarrow e\downarrow S\uparrow}(k)|^2\right)\right]
\left(\frac{\partial n_{k\uparrow}}{\partial\mu}\right)\\
\nonumber
+
\left[
P_{S\downarrow}|A_{e\downarrow S\downarrow\rightarrow e\downarrow S\downarrow}(k)|^2
+P_{S\uparrow}
\left(
|A_{e\downarrow S\uparrow\rightarrow e\downarrow S\uparrow}(k)|^2+
\right.\right.
%\\
%\nonumber
\left.\left.
|A_{e\downarrow S\uparrow\rightarrow e\uparrow S\downarrow}(k)|^2
\right)
\right]
\left(\frac{\partial n_{k\downarrow}}{\partial\mu}\right)\\
\nonumber
+
\left[
P_{S\downarrow}
|A_{e\uparrow S\downarrow\rightarrow e\downarrow S\uparrow}(k)|^2
-P_{S\uparrow}
|A_{e\downarrow S\uparrow\rightarrow e\uparrow S\downarrow}(k)|^2
\right]
%\\
%\times
\left[
 n_{k\uparrow}(\mu)\left(\frac{\partial n_{k\downarrow}}{\partial\mu}\right)
-n_{k\downarrow}(\mu)\left(\frac{\partial n_{k\uparrow}}{\partial\mu}\right)
\right]
kdk
\}.
\label{EqCond}
\end{eqnarray}
\end{widetext}
In this formula we have introduced a number of parameters. The mean number of electrons with a definite spin projection and wavevector $n_{k\uparrow,\downarrow}$ is the same if we do not account for the Zeeman effect and reads:
\begin{equation}
n_{k\uparrow,\downarrow}=\frac{1}{\mathrm{e}^{\frac{E_k-\mu}{k_BT}}+1}.
\end{equation}
Here $\mu$ is the chemical potential in the quantum wire. It is shifted by the value $eV_{d}$ in the right lead; $P_{S\uparrow}$ and $P_{S\downarrow}$ are probabilities of the localized electron findings in spin-up and spin-down states, given by the formulae:	 
\begin{eqnarray}
P_{S\uparrow}=\frac{\mathrm{e}^{g\mu_BB/k_BT}}{\mathrm{e}^{g\mu_BB/k_BT}+\mathrm{e}^{-g\mu_BB/k_BT}};\\
P_{S\downarrow}=\frac{\mathrm{e}^{-g\mu_BB/k_BT}}{\mathrm{e}^{g\mu_BB/k_BT}+\mathrm{e}^{-g\mu_BB/k_BT}},
\end{eqnarray}
and in the case of absent Zeeman splitting, $P_{S\uparrow}=P_{S\downarrow}=1/2$.

%-----Scattering matrix

Parameters $A$ (namely, $A_{e\uparrow S\uparrow\rightarrow e\uparrow S\uparrow}$, $A_{e\uparrow S\downarrow\rightarrow e\uparrow S\downarrow}$, $A_{e\uparrow S\downarrow\rightarrow e\downarrow S\uparrow}$; $A_{e\downarrow S\downarrow\rightarrow e\downarrow S\downarrow}$, $A_{e\downarrow S\uparrow\rightarrow e\downarrow S\uparrow}$, $A_{e\downarrow S\uparrow\rightarrow e\uparrow S\downarrow}$) are the amplitudes of transmission through the AB ring; the subscripts denote the spin states before and after the scattering event. To find the transmittion amplitudes (using other scattering amplitudes), we will address the scattering matrix approach discussed in works. \cite{ShelykhPRB2005,Shelykh2007}

The amplitudes of the transmitted and reflected waves can be found from the conditions of the conservation of the flux at the contacts connecting the AB ring and the leads. As we account for the exchange interaction inside the ring, we should consider separately the cases when the spins of the propagating and localized electrons are aligned parallel or anti- parallel. In the further discussion we present the equations for the simplest case of the localized spin $J=1/2$ only. The generalization for the case $J>1/2$ is straightforward and can be done according to the lines presented in Ref.\onlinecite{ShelykhPRB2005}.

First, let us consider the case in which the spins of both the propagating and the localised electrons are parallel and thus electrons are in the triplet configuration. Note, that as exchange interaction conserves the total spin projection of the interacting electrons, in this case the spin- flip processes in the ring are impossible. Moreover, as we consider the case of spin- conserving contacts between the ring and the leads, the orientations $\uparrow\uparrow$ and $\downarrow\downarrow$ can be treated separately. The problem thus effectively reduces to one for the transport of the spinless particles.
Introducing the spin-independent scattering matrix for the contacts $\hat{S}$  one can obtain the following system of linear algebraic equations connecting the transmission and reflection amplitudes, $A=A_{e\uparrow S\uparrow\rightarrow e\uparrow S\uparrow}=A_{e\downarrow S\downarrow\rightarrow e\downarrow S\downarrow}$ and $B=B_{e\uparrow S\uparrow\rightarrow e\uparrow S\uparrow}=B_{e\downarrow S\downarrow\rightarrow e\downarrow S\downarrow}$ with the amplitudes of the waves propagating in the ring  $b_j$,$c_j$ ($j=+,-$) and shown in Fig.~1:
\begin{eqnarray}
\nonumber
\left( \begin{array}{c} b_- \\ A \\ c_+ \end{array} \right)
&=&
\hat{S}
\cdot
\left( \begin{array}{c} b_+\tau_t^+ \\ 0 \\ c_-\tau_t^- \end{array} \right)
=
\left(
\begin{array}{ccc}
r & \varepsilon & t  \\
\varepsilon & \sigma & \varepsilon  \\
t & \varepsilon & r
\end{array}
\right)
\cdot
\left( \begin{array}{c} b_+\tau_t^+ \\ 0 \\ c_-\tau_t^- \end{array} \right);\\
\label{EqScatMatr1}
\left( \begin{array}{c} b_+ \\ B \\ c_- \end{array} \right)
&=&
\hat{S}
\cdot
\left( \begin{array}{c} b_-\tau_t^- \\ 1 \\ c_+\tau_t^+ \end{array} \right).
\end{eqnarray}
The QPCs connecting the ingoing and outgoing leads to the ring are considered to be identical. The scattering matrix is characterized by the parameters $r,t,\epsilon,\sigma$ which have the following physical meaning. The parameters $r$ and $t$ are reflection and transmission amplitudes of the QPCs inside the AB ring; $\sigma$ is the reflection amplitude from the lead to itself and $\varepsilon$ is the transmission amplitude from a lead to the AB ring or from the AB ring to a lead. The scattering amplitudes $r$, $t$, $\sigma$, and $\varepsilon$ are assumed to be real numbers. These parameters depend on the properties of the junction, in particular on the band mismatch between the leads and the AB ring which can be electrically induced by the vertical gate voltage in the ring region. The condition of flux conservation resulting in the hermicity of the scattering matrix allows to reduce the number of its independent elements. According to Buttiker~\cite{Buttiker1984,Endquist1981,Taniguchi1999}, the following parametrization can be used:
\begin{eqnarray}
r=\frac{\lambda_1+\lambda_2\sqrt{1-2\varepsilon^2}}{2};\\
\nonumber
t=\frac{-\lambda_1+\lambda_2\sqrt{1-2\varepsilon^2}}{2};\\
\nonumber
\sigma=\lambda_2\sqrt{1-2\varepsilon^2},
\end{eqnarray}
where $\lambda_{1,2}=\pm 1$. Therefore, the effect of the QPCs on the scattering of a particle in the AB ring is defined by single parameter: $\varepsilon\in[-1/\sqrt{2},1/\sqrt{2}]$. The case $\varepsilon=1/\sqrt{2}$ corresponds to the fully transmitting contact, the case $\varepsilon=0$ to the fully reflecting one.

The terms $\tau_t^{\pm}$ are the phase shifts between the clockwise and anticlockwise traveling electron waves, correspondingly:
\begin{equation}
\tau_t^{\pm}=exp\left[{i\left(\pi k_t R\pm \frac{e\Phi}{2\hbar}\right)}\right],
\end{equation}
where $\Phi=\pi R^2B$ is the magnetic flux through the ring, $R$ is the radius of the AB ring, and $e$ is the electron charge. The wavevector $k_t$ correspond to triplet configuration of the interacting spins and is defined by Eq.~(\ref{Eqk1}). In the absence of an external magnetic field, the phase shift is equal for the electrons moving in both the clockwise and anticlockwise directions.

%------

Now, let us consider the situation when the spins of the localised and propagating electrons are untiparallel, and the spin flip process can occur. For instance, consider the case when propagating electron is initially in the spin-up state and localized electron is in spin-down state (the opposite case is fully equivalent). The amplitudes of scattering then become spin dependent and instead of the single transmission amplitude one needs to introduce two amplitudes corresponding to the cases of the conservation of the spin of the propagating electron and its spin flip, $A_{e\uparrow S\downarrow\rightarrow e\uparrow S\downarrow}$ and $A_{e\uparrow S\downarrow\rightarrow e\downarrow S\uparrow}$ respectively (the same is valid for the amplitudes of reflection $B_{e\uparrow S\downarrow\rightarrow e\uparrow S\downarrow}$ and $B_{e\uparrow S\downarrow\rightarrow e\downarrow S\uparrow}$). Therefore, the system of the equations for the amplitudes becomes more complicated and reads:
\begin{eqnarray}
\label{EqScatMatr2}
\left(
\begin{array}{c} b_{t-} \\ A_{e\uparrow S\downarrow\rightarrow e\uparrow S\downarrow} \\ c_{t+} \\
b_{s-} \\ A_{e\uparrow S\downarrow\rightarrow e\downarrow S\uparrow} \\ c_{s+}
\end{array}
\right)
=
\left(
\begin{array}{ccc}
\hat{S} & 0 \\
0 & \hat{S}
\end{array}
\right)
\cdot
\left(
\begin{array}{c}
b_{t+}\tau_{t}^+ \\ 0 \\ c_{t-}\tau_{t}^-
\\
b_{s+}\tau_{s}^+ \\ 1 \\ c_{s-}\tau_{s}^-
\end{array}
\right);
\end{eqnarray}
%
%
%-----
%
%
\begin{eqnarray}
\nonumber
\left(
\begin{array}{c}
b_{t+} \\ B_{e\uparrow S\downarrow\rightarrow e\uparrow S\downarrow} \\ c_{t-} \\
b_{s+} \\ B_{e\uparrow S\downarrow\rightarrow e\downarrow S\uparrow} \\ c_{s-}

\end{array} \right)
=
\left(
\begin{array}{ccc}
\hat{S} & 0 \\
0 & \hat{S}
\end{array}
\right)
\cdot
\left(
\begin{array}{c}
b_{t-}\tau_{t}^- \\ 1 \\ c_{t+}\tau_{t}^+  \\
b_{s-}\tau_{s}^- \\ 0 \\ c_{s+}\tau_{s}^+
\end{array}
\right),
\end{eqnarray}
where
\begin{equation}
\tau_{t,s}^{\pm}=exp\left[{i\left(\pi k_{t,s} R\pm \frac{e\Phi}{2\hbar}\right)}\right],
\end{equation}
and $k_{t,s}$ are given by Eqs.~(\ref{Eqk1}), (\ref{Eqk2}). It should be noted, that as initial configuration $|e\uparrow S\downarrow\rangle$ is not an eigenstate of the exchange Hamiltonian~(\ref{EgDirExHam}), both singlet and triplet configurations corresponding to different wavenumbers of the propagating electrons are possible inside the ring.

Together, Eqs.~(\ref{EqCond}), (\ref{EqScatMatr1}) and (\ref{EqScatMatr2}) allow for calculation the conductance in the ballistic regime.

%-----------------------------------------------------------------
%-----------------------------------------------------------------
%-----------------------------------------------------------------

\section{Results and discussion}

In the calculations presented herein below we used the following set of parameters: $m=0.063$ $m_e$, $V_{dir} =0.07e^2/t(\pi\epsilon_0 R)$, $V_{ex} =\pm 0.5V_{dir}$. The radius of the ring was taken equal to $R = 100$ $nm$. We also considered the cases of different values of the uncompensated spin in the ring $J$ ($J=1/2,3/2,5/2,7/2$). The temperature was put zero. Results of modeling are presented in Figs.~2-5.

Figures~2 illustrate the influence of the many-body interaction on the AB ring conductance for the cases of zero and non-zero magnetic fields.
\begin{figure}[ht]
\centering
\subfigure[Ferromagnetic interaction ($V_{ex}>0$)]{
   \includegraphics[width=0.9\linewidth] {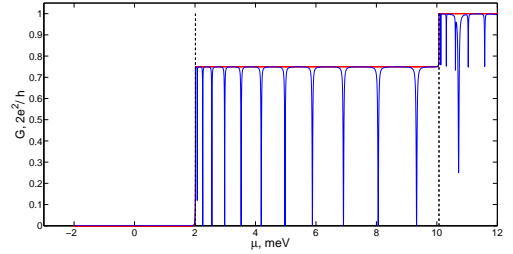}
   %\label{FigGmu:a}
 }\\
 \subfigure[Antiferromagnetic interaction ($V_{ex}<0$)]{
   \includegraphics[width=0.9\linewidth] {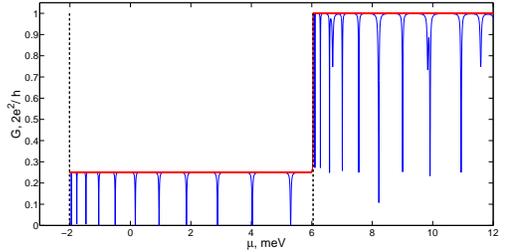}
   %\label{FigGmu:b}
 }
\label{FigGmu}
\caption{The dependence of the conductance of the AB with single localized spin on the Fermi energy in the case of (a) ferromagnetic  ($V_{ex}>0$) and (b) antiferromagnetic ($V_{ex}<0$) interaction between the spins of localized and ballistic electrons. The contacts connecting AB ring to the leads are considered to be fully transparent, $\varepsilon=1/\sqrt{2}$. Red curves correspond to the case of zero magnetic field ($B=0$), blue curves to the case $B=1.7$ T. For $B=0$ the conductance reveal the same fractional quantization pattern as those of the individual QPC with localized spin. Magnetic field introduces additional AB phaseshifts, and conductance reveals oscillatory pattern.}
\end{figure}

Both direct and exchange interactions of the moving and localized electrons form an effective potential barrier for the carriers entering into the ring, thus determining the conductance.
In general, there are two such sub-barriers in the ring region, for triplet and singlet spin orientations corresponding to the energies: $V_{dir}-V_{ex}J$ and $V_{dir}+3V_{ex}$.

Figure~2a corresponds to the case of ferromagnetic interaction ($V_{ex}>0$).
The behavior of $G$ has a clear explanation. Let us first consider the case when magnetic field is absent shown by the red line. The conductance $G$ is zero below the $\mu=V_{dir}-V_{ex}J$ (the lowest step of the barrier) since independently on the mutual orientation of the spins of propagating and localized electrons the energy of a particle is not enough to overcome even the lowest barrier and all carriers are reflected.
Further, if the chemical potential lies in the range $[V_{dir}-V_{ex};V_{dir}+3V_{ex}]$ shown by two vertical liens on the plot the carriers in triplet configuration can enter the ring while the carriers in the singlet configuration can not. As there are three states corresponding to the triplet configuration and one state corresponding to the singlet configuration and contacts are considered to be fully transparent, the conductance in these regime equals to $G=0.75(2e^2/h)$. If $\mu>V_{dir}+3V_{ex}J$, the electrons in both singlet and triplet configurations can enter the ring and conductance reaches the value of the elementary conductance quantum, $G=2e^2/h$. The conductance pattern is thus equivalent to those revealed by an individual QPC with localized spin studied in Refs.\cite{Rejec2000,Rejec2003,Shelykh2006}.
\begin{figure}[ht]
\label{FigGB}
\centering
\subfigure[Ferromagnetic interaction ($V_{ex}>0$)]{
   \includegraphics[width=0.9\linewidth] {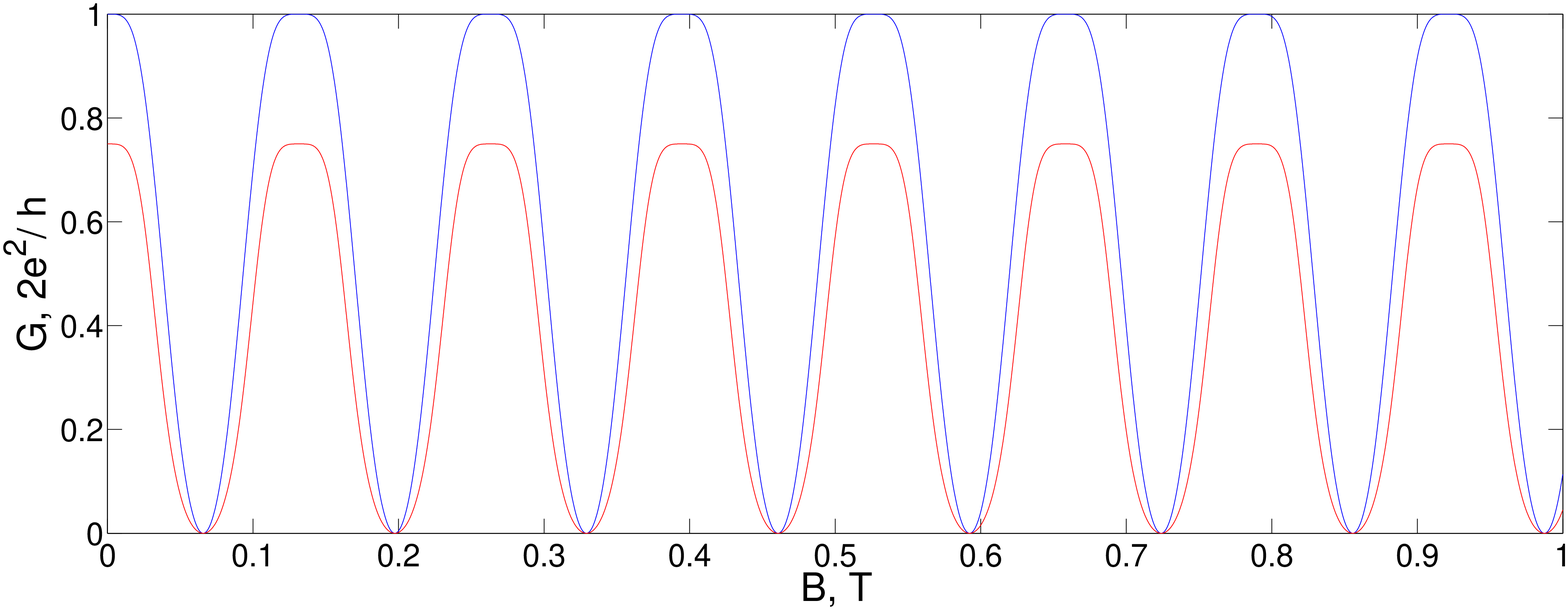}
   %\label{fig3:subfig1}
 }
 \subfigure[Antiferromagnetic interaction ($V_{ex}<0$)]{
   \includegraphics[width=0.9\linewidth] {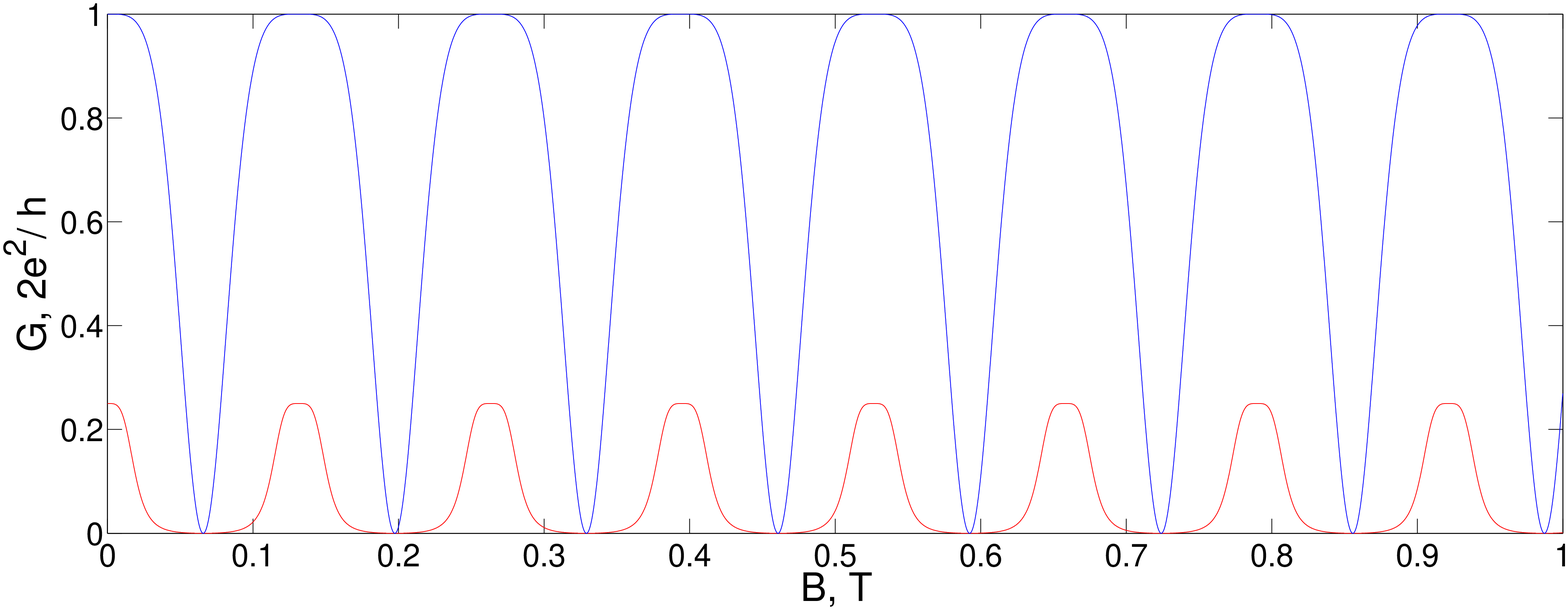}
   %\label{fig3:subfig2}
 }
\caption{Conductance of the AB ring with exchange interaction vs magnetic field for different Fermi energies: $5.0$ (red curves) and $15.0$ $meV$ (blue curves) in the case of (a) ferromagnetic interaction ($V_{ex}>0$) and (b) antiferromagnetic interaction ($V_{ex}<0$). The value $\mu=5$ meV corresponds to the second steps on the $G(\mu)$ staircases (see Fig.~2a,b). Therefore, the maxima of the amplitudes of oscillations correspond to (a) $0.75(2e^2/\hbar)$ and (b) $0.25(2e^2/\hbar)$.}
\end{figure}

If one applies an external magnetic field inducing different AB phaseshifts for the electrons moving clockwise and anticlockwise and interference is not always constructive, the conductance as a function of the chemical potential reveals oscillations instead of the plateaus. The amplitude of these oscillations is $0.75(2e^2/h)$ if chemical potential lies in the range $[V_{dir}-V_{ex};V_{dir}+3V_{ex}]$ and only triplets are transmitted and increases to $2e^2/h$ if chemical potential is increased above the value $V_{dir}+3V_{ex}J$ if transport of singlets also becomes possible.
\begin{figure}[ht]
\label{FigGeps}
\centering
\subfigure[$\varepsilon= 0.90\times1/\sqrt{2}$]{
   \includegraphics[width=0.99\linewidth] {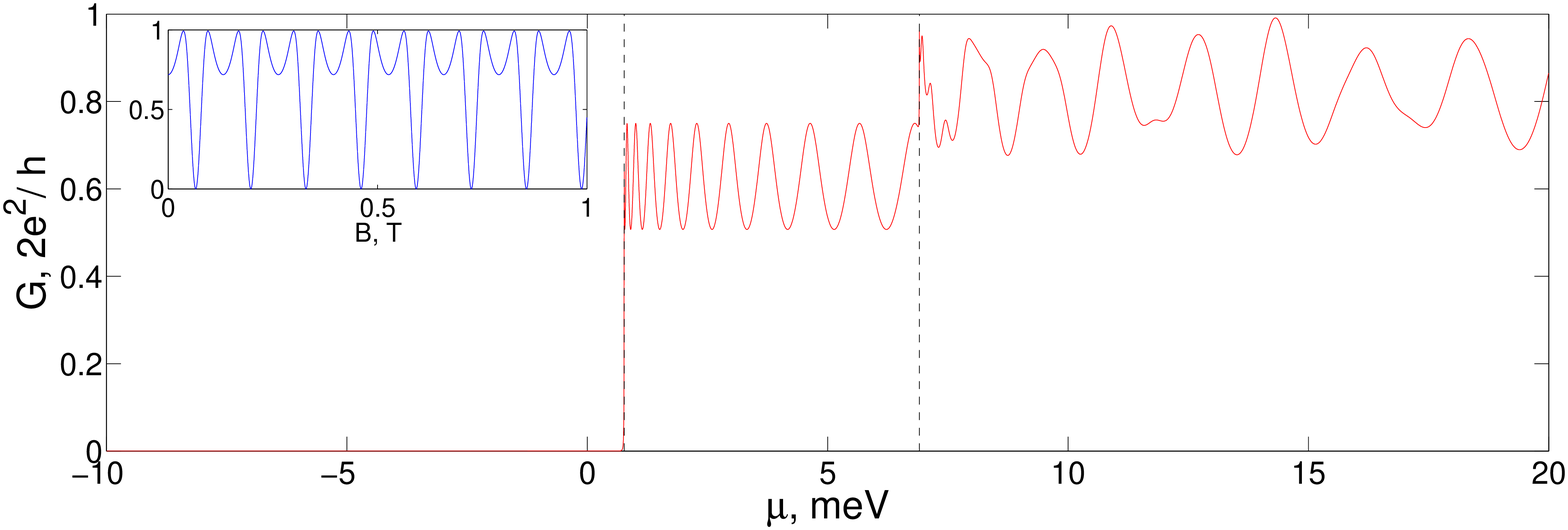}
   %\label{FigGeps:a}
 }
 \subfigure[$\varepsilon= 0.35\times1/\sqrt{2}$]{
   \includegraphics[width=0.99\linewidth] {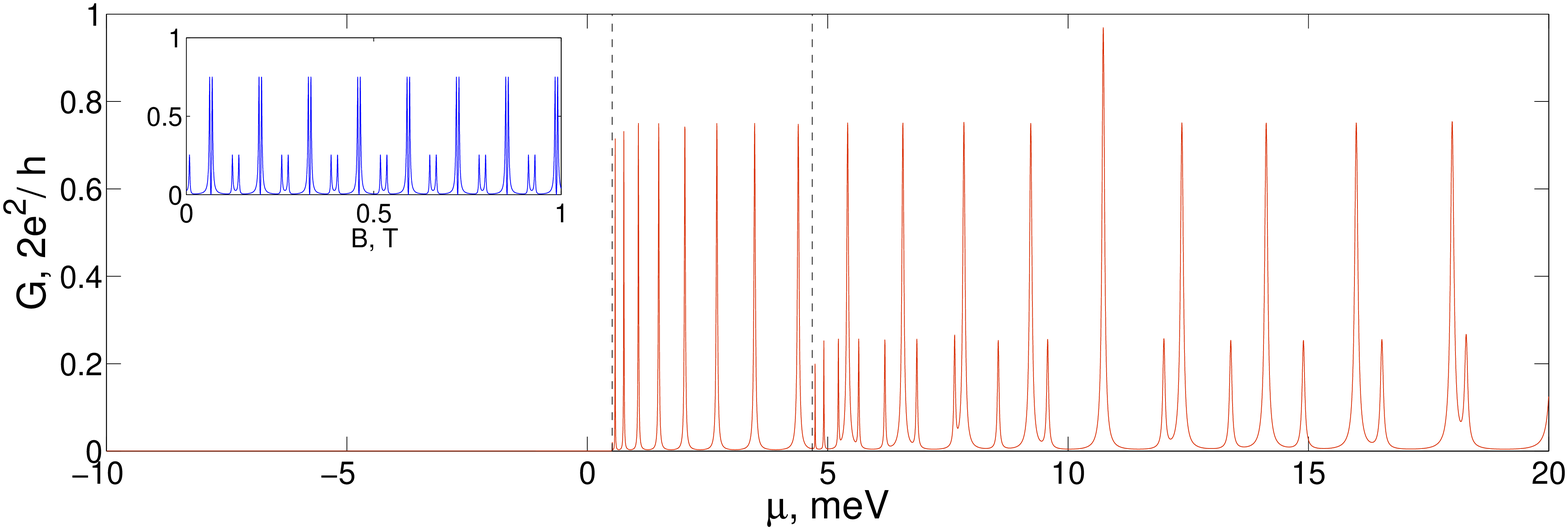}
   %\label{FigGeps:b}
 }
\caption{The influence of the transparency of the contacts on the conductance pattern of the ring with ferromagnetic exchange interaction. Two cases are considered: (a) $\varepsilon=0.90\times 1/\sqrt{2}$ and (b) $\varepsilon=0.35\times 1/\sqrt{2}$. Main plots depict the dependence $G(\mu)$ for $B=0$; Insets show the dependencied of the conductance on magnetif field for $\mu=15$ meV.}
\end{figure}

Figure~2b corresponds to the case of antiferromanetic interaction ($V_{ex}<0$). Now, the singlet configuration becomes energetically preferable. The conductance $G$ is zero if chemical potential lies below the $\mu=V_{dir}+3V_{ex}$
Further, if chemical potential lies in the region $[V_{dir}+3V_{ex};V_{dir}-V_{ex}]$ the conductance becomes equal to $0.25(2e^2/\hbar)$ since only electrons in singlet configuration can enter the ring. Finally, at $\mu>V_{dir}-V_{ex}$ the ring becomes transparent for any spin orientation and the value $G=2e^2/\hbar$ is recovered.
\begin{figure}[ht]
\label{FigGJ}
\centering
\subfigure[Ferromagnetic interaction ($V_{ex}>0$)]{
   \includegraphics[width=0.99\linewidth] {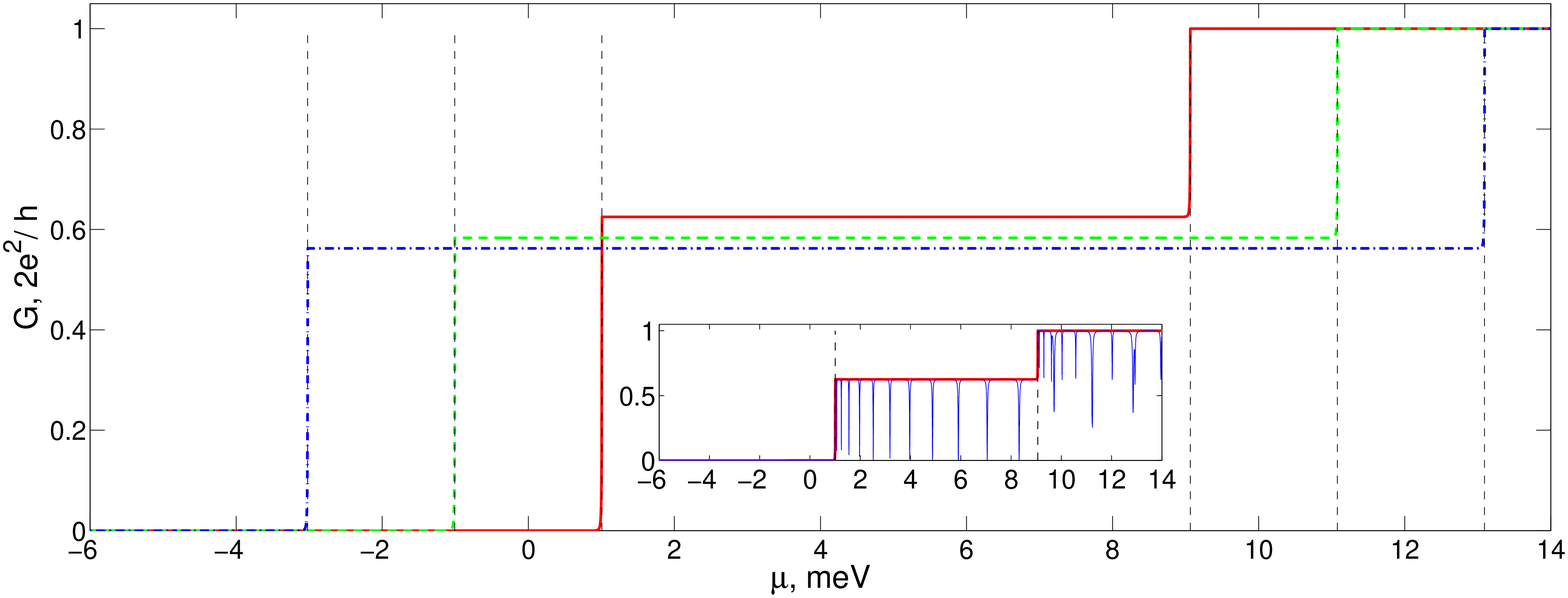}
   %\label{FigGJ:a}
 }
 \subfigure[Antiferromagnetic interaction ($V_{ex}<0$)]{
   \includegraphics[width=0.99\linewidth] {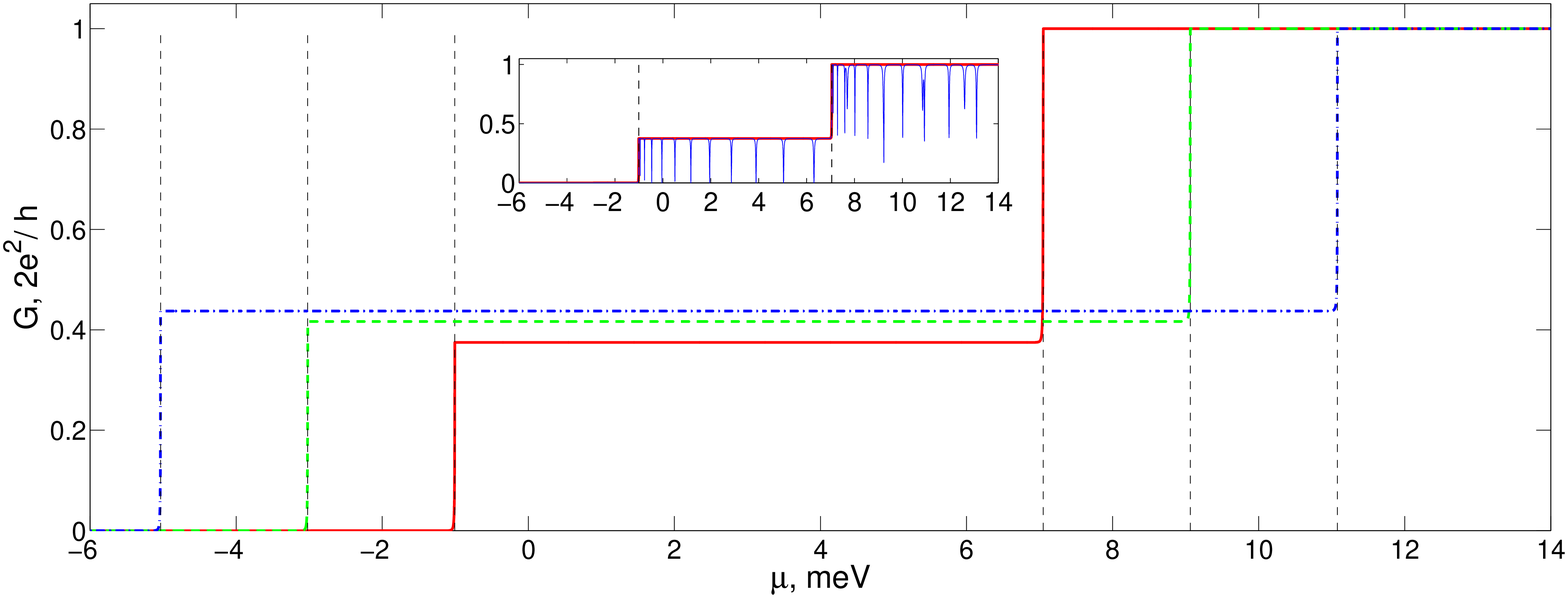}
   %\label{FigGJ:b}
 }
\caption{Conductance of the AB ring dependence on the Fermi energy for different values of the uncompensated spin localized in the ring $J$: $J=3/2$ (red), $J=5/2$ (green) and $J=7/2$ (blue) for the cases of (a) ferromagnetic and (b) antiferromagnetic interaction. Main plots correspond to zero magnetic field. Insets illustrate the magnetic field dependence: $B=0$ (red) and $B=1.7$ T (blue) for the case $J=3/2$. The transparency parameter $\varepsilon= 1/\sqrt{2}$. }
\end{figure}

Figures~3 illustrate the dependence of the conductance on the external magnetic field for different values of the chemical potential and $\varepsilon=1/\sqrt{2}$. An oscillatory behavior revealing AB effect is observed. This happens since the conductance of the ring is governed by the phase factors which become different for the clockwise ($\tau^+$) and anticlockwise ($\tau^-$) propagating waves. It should be noted, that Figs.~3a,b are in good correspondence with Figs.~2a,b.
The amplitude of the AB oscillations is defined by the value of the chemical potential. It changes from $0.75(2e^2/h)$ to $2e^2/h$ when $\mu$ increases above $V_{dir}+3V_{ex}$ for the ferromagnetic case and changes from $0.25(2e^2/h)$ to $2e^2/h$ when $\mu$ increases above $V_{dir}-V_{ex}$ for the antiferromagnetic case.
We have also analyzed the change of the conductance pattern if transparency of the contacts described by the parameter $\varepsilon$ is changed. The results are shown in Figs.~4 for ferromagnetic case only (aniferromagnetic case is similar). The deviation from the case of fully transparent contacts corresponding to $\varepsilon=1/\sqrt{2}$ is revealed by the onset of the oscillatory pattern in the dependence of the conductance on the chemical potential even at zero magnetic field. If $\mu\in[V_{dir}+3V_{ex};V_{dir}-V_{ex}]$, then only electrons in the triplet configuration can pass and the amplitude of the oscillations is $0.75(2e^2/h)$. When $\mu$ exceeds the value of $V_{dir}-V_{ex}$, the ring becomes accessible for both singlet and triplet configurations.
The values of the wavenumber for the ballistic electron and thus conditions of the constructive interference in these two configurations are different. If the deviation from the condition of full transparency is small (e.g. $\varepsilon=0.9/\sqrt{2}$), this results in the onset of the broad oscillations with the amplitude a bit smaller then $2e^2/h$ in the region $\mu>V_{dir}-V_{ex}$, as shown in Fig. 4a. However, if one decreases the transparency further, the resonances become sharper, and conductance pattern at $\mu>V_{dir}-V_{ex}$ consists of the series of well resolved peaks of the height $0.75(2e^2/h)$ and $0.25(2e^2/h)$ corresponding to resonant transmission of triplets and singlets, respectively.

In the dependence of the conductance on magnetic field shown at the insets of the figures, the decrease of the transparency of the contacts leads to the appearance of the higher harmonics connected with the increased probabilities of the round trips inside the ring and characteristic to the transition from Aharonov- Bohm to Aharonov-Altshuler-Spivak oscillations, typical for the weak localization regime. \cite{AAS,ShelykhPRB2005}
%

%-----Fig 5
Finally, we have analyzed how the increase of the localized spin affects the conductance patterns of the AB rings with exchange interaction. The results are shown in Figs.~5. In the case of transparent contacts, the increase of the spin leads to the decrease of the value of the sub- step from $0.75(2e^2/h)$ to $[(J+1)/(2J+1)](2e^2/h)$ in case of the ferromagnetic interaction and increase from $0.25(2e^2/h)$ to $[J/(2J+1)](2e^2/h)$ for antiferromagnetic interaction. This is in agreement with the picture of the fractional quantization of the ballistic conductance presented in Ref.\cite{Shelykh2006}. The application of the external magnetic field leads to the onset of the oscillatory behavior, similar to those observed for $J=1/2$.

\section{Conclusions}
In conclusion, we have considered ballistic conductance of an Aharonov-Bohm ring containing localized uncompensated spin accounting for the exchange interaction between the spins of propagating and localized electrons. We have shown that exchange interaction drastically modifies the conductance as function of magnetic field and chemical potential. The obtained results are in agreement with the concept of the fractional quantization of the ballistic conductance proposed earlier.\\
%__________________________________________________________________________
%__________________________________________________________________________
%__________________________________________________________________________

\section*{ACKNOWLEDGEMENTS}
The work was supported by FP7 IRSES project ``SPINMET''.
R.G.P. thanks the University of Iceland for hospitality.
I.G.S. acknowledges support of the Eimskip Fund.

%\section*{References}

%*****************************
%*****************************
%*****************************

\end{document}